\documentclass[12pt, a4paper]{article}
\usepackage{geometry}
\usepackage[utf8]{inputenc}

\usepackage[T1]{fontenc} 
\usepackage{graphicx,epstopdf} 

\usepackage{bm}
\usepackage{secdot}
\usepackage{mathptmx}
\usepackage{float}
\usepackage{soul}
\usepackage{authblk}

\RequirePackage{graphicx}
\RequirePackage{amsmath}
\RequirePackage{amssymb}

\title{A Viable Dark Fluid Model}

\author{Esraa Elkhateeb} 
\affil{\small Physics Department, Faculty of Science, Ain Shams University., Abbassia 11566, Cairo, Egypt}
\author{\\ \footnotesize{}}

\begin{document}
\date{}

\maketitle

\begin{abstract}
We consider a cosmological model based on a generalization of the equation of state proposed by Nojiri and Odintsov \cite{noj} and $\check{\texttt{S}}$tefan$\check{\texttt{c}}$i$\acute{\texttt{c}}$ \cite{stef}, \cite{stef1}. We argue that this model works as a dark fluid model which can interpolate between dust equation of state and the dark energy equation of state. We show how the asymptotic behavior of the equation of state constrained the parameters of the model. The causality condition for the model is also studied to constrain the parameters and the fixed points are tested to determine different solution classes. Observations of Hubble diagram of SNe Ia supernovae are used to further constrain the model. We present an exact solution of the model and calculate the luminosity distance and the energy density evolution. We also calculate the deceleration parameter to test the state of the universe expansion. 
\end{abstract}


\section{Introduction}
Recent cosmological observations favoring the scenario of spatially flat and accelerated universe. Observations come from Hubble diagram of SNe Ia supernovae are best fitted by spatially flat accelerated cosmological models \cite{per} - \cite{teg}. Besides, other observations from Cosmic Microwave Background Radiation (CMBR) and Baryon Acoustic Oscillation (BAO) are also strong evidences for this scenario \cite{bran} - \cite{rie}. This currently observed speeding up expansion of the universe leads to the fact that there is a repulsive force acting in the universe space-time.

According to our current understanding of the universe history, the radiation era was followed by a matter dominated era during which most of the universe structures like stars, galaxies and galaxy clusters we now observe were formed by gravitational instabilities. Before these ordinary matter structures were formed, a particular form of gravitating but non-baryonic sort of matter structures was formed. This sort of matter which is non-relativistic does not interact with radiation, and so is called Cold Dark Matter (CDM). The perturbations of this dark matter (DM) collapse first because its particles don't couple to photons as they don't interact with them. Their gravitational potentials then guide for baryon collapse \cite{jaf} - \cite{low}. Late in this matter dominated phase, a new but very peculiar form of invisible and non-gravitating "matter" started to dominate. This sort of "matter" does not interact with baryons, radiations, or any other sort of visible matter. It is now known as Dark Energy (DE).

Observational evidences show that, see for ex. \cite{per} and \cite{ries1} - \cite{tonr}, our universe at present contains approximately $4\%$ only of radiation and baryonic matter that is well known by the standard model of particle physics and can be detected in laboratory and so is considered visible. While about $26\%$ is the non-baryonic dark matter. The rest of our universe content, which is about $70\%$ is the exotic component known as dark energy. As a result, this dominated dark energy component antigravitates leading to the present observed universe expansion.
 
Although dark energy is an acceptable and effective description for such large scale homogeneous and isotropic accelerated universe,  identifying its origin and nature is one of the most debated questions in physics and cosmology. Large variety of models are proposed to describe the nature of such dark energy \cite{yue}, \cite{cop}. As the energy density due to DE remains unchanged throughout the universe space-time, the simplest choice was to revive the idea of Einstein about the cosmological constant $\Lambda$ \cite{pee}, \cite{wie}. This together with Cold Dark Matter (CDM) provides the standard model for cosmology $\Lambda$CDM \cite{wie} - \cite{pad}, which appears to be in a very good agreement with observational data. In Einstein's field equation, this term has the concept of intrinsic energy density of vacuum \cite{carr1}. However, this arises an important question, why the observed value of vacuum energy density is very far below that is predicted from particle physics?. This is the famous fine tuning problem. Another important question is why the energy densities of the vacuum and the matter are of the same order of magnitude today?, the famous coincidence problem.

Dynamical models are other alternative to the cosmological constant. Some of these are the minimal coupled scalar fields or quintessence \cite{weh} - \cite{fol}, k-essence \cite{rub} - \cite{mats}, Chaplygin gas \cite{baf} - \cite{din}, all have an equation of state parameter $\omega\geq-1$. These models adopt the idea of the possibility of unification of the two dark sectors, dark energy and dark matter. Models with $\omega\prec-1$ knowing as phantom energy models \cite{lud}, \cite{odn1} consider a growing vacuum energy with expansion. A more general class of models considers an equation of state parameter varies with the scale factor, or the red shift \cite{pee}, \cite{sah}, \cite{pad}, \cite{alc}, \cite{eli}. There are also some observational analysis which consider the variation of $\omega$ with the red shift and with time \cite{ala}, \cite{ale}. Modification of the governing equation, or the so called modified gravity theory \cite{bamb}, \cite{moda}, is also one way of thinking (For a general review of different DE models see \cite{odn3}).

The idea of unification of DE and DM is a promising one and Chaplygin gas model is one of the most acclaimed models representing it. This model was proposed by Kamenshchik et. al. as early as 2001 \cite{kam}  where they assumed that the universe, within the framework of standard cosmology, is filled with an exotic fluid obeys the Chaplygin gas equation of state (EoS)
\begin{equation} 
p=-A / \rho^\alpha
\label{cha}
\end{equation}
where $A$ is a positive constant and $\alpha=1$. This model has the advantage of smooth transition between different phases of the universe. Many authors then extended this model using the idea of generalizing or modifying the EoS of the background fluid in order to improve the behavior of the fluid throughout the universe evolution. In 2002, Bento et. al. \cite{bent} extended the model to a more general form known as the generalized Chaplygin gas (GCG) model where the $\alpha$ constant is considered in the range $0 \prec \alpha \leq 1$. Other various models based on Chaplygin gas were also proposed such as the modified Chaplygin gas model introduced in 2002 by Benaoum \cite{bena} where he could interpolate between standard fluids at high energy densities and Chaplygin gas fluids at low energy densities using the EoS
\begin{equation} 
p= A \rho - \frac{B}{\rho^\alpha}
\label{cha2}
\end{equation}
where $\alpha \geq 1$ and $A$ and $B$ are positive constants, and the hybrid Chaplygin gas model introduced by Bili$\acute{\texttt{c}}$ et. al. in 2005 \cite{bili} and leads to transient acceleration.

In this work, we have adopted a class of dynamical models in which the EoS has the advantage of interpolation between two different Chaplygin Gas models. Our fluid is a barotropic one with an EoS that can be considered as a correction to the vacuum EoS by a one asymptotes between two power laws, so that it has the advantage of interpolation between the two equations of state of the DE and DM, which might describe some sort of smooth phase transition. Another advantage of this model is that it is not only interpolating between dust and DE in early and late times, but also it has a more general EoS for DE that enables the cosmological constant as a special case. This has the more advantage of a general asymptotic solution than Chaplygin gas which goes to the cosmological constant. We present an exact solution of the equations which enables more accurate results for the calculations of the luminosity distance, while observations of Hubble diagram of SNe Ia supernovae are used to further constrain the model parameters. We study the energy density and the equation of state parameter evolutions. The analysis of the EoS shows that the dynamical properties of both dark sectors can be described through this model which lets our choice seems natural because of the smooth transition. We also calculate the deceleration parameter to examine the expansion of the universe due to the model.
          
The rest of the article is organized as follows; in the following section we have formulated the model. In section \ref{sectr} the mathematical treatment of the model is made. In section \ref{anal} we study the phase diagram of the model and its classes of solutions. Section \ref{case1} treats the first case of study of our model where the causality constraints are examined and the dynamics of the universe is studied where the equation of the luminosity distance as a function of red shift is solved and parameters are constrained through Hubble diagram observations, the energy density and the deceleration parameter evolutions due to the model are examined. Section \ref{case2} treats the second case of study of our model. Section \ref{seconc} concludes our work.

\section{Building the Model} \label{secmod}
We are working through the standard FRW cosmology where the metric in the spatially flat geometry is
\begin{equation}
ds^2=dt^2-a^2(t)(dr^2+r^2d\Omega^2)
\label{metric}
\end{equation}
where we consider units with $c=1$. Using the energy momentum tensor given by 
\begin{equation}
T_{\mu\nu}=(\rho+P)U_\mu U_\nu-P g_{\mu\nu}
\label{tensor}
\end{equation}
where in comoving coordinates $U_\mu={\delta^0}_\mu$, the Einstein's equation will lead to the Friedman equations
\begin{align}
	\frac{\dot{a}^2}{a^2}=\frac{8 \pi G}{3}\rho \label{dot}\\
	\frac{\ddot{a}}{a}=-\frac{4 \pi G}{3}(\rho+3P) \label{ddot}
\end{align}
while the conservation equation ${{T_\nu}^\mu};\mu=0$ gives
\begin{equation}
\dot{\rho}=-3 \frac{\dot{a}}{a} (\rho + P)
\label{cons}
\end{equation}
Now we have to consider an EoS to close the system. It is a fact that dark energy dominates the universe today and an adequate EoS for this component is $p=\omega \rho$. Observations indicate that the EoS parameter $\omega$ is close to $-1$ and people are now targeting to confirm whether $\omega$ is constant or varies as the universe expands. Owing to this fact, Nojiri and Odintsov \cite{noj} considered a general EoS of the form
\begin{equation} 
p= -\rho - f(\rho)
\label{cha3}
\end{equation}
where $f(\rho)$ is an arbitrary function may be considered as a correction to the standard DE equation of state. $\check{\texttt{S}}$tefan$\check{\texttt{c}}$i$\acute{\texttt{c}}$ \cite{stef}, \cite{stef1} studied in details the case of $f(\rho) \propto \rho^\alpha$ with a constant $\alpha \neq 1$, and Nojiri and Odintsov \cite{noj1}, \cite{odn4} studied the future singularity associated with this model.

In this work we adopted a barotropic fluid with an EoS which is also a correction to the standard dark energy EoS. The fluid pressure has a general form for the density dependence
\begin{equation}
P=-\rho + \frac{\gamma \rho^n}{1+\delta \rho^m}
\label{van1}
\end{equation}
Where $\gamma$, $\delta$, $n$, and $m$ are constants which are considered as free parameters. This enables interpolation between different powers for the density, so that the phase transitions during the universe's evolution are described smoothly. The EoS parameter is given by
\begin{equation}
\omega=-1 + \frac{\gamma \rho^{n-1}}{1+\delta \rho^m}
\label{par1}
\end{equation}

\section{The model} \label{sectr}
Let's now study the evolution of the physical quantities of the universe due to our model. Considering units with $8 \pi G=1$, eqns (\ref{dot})-(\ref{cons}) reduce to

\begin{align}
H^2=\frac{1}{3} \rho	\label{hsq} \\
\frac{\ddot{a}}{a}=H^2+\dot{H}=-\frac{1}{6}(\rho+3P) \label{hddot}  \\
\dot{H}=-\frac{1}{2} (\rho + P)  \label{rodot}
\end{align}
where $H=\frac{\dot{a}}{a}$ is the Hubble parameter. Combining eq(\ref{van1}) to eq(\ref{rodot}) we get
\begin{equation}
\dot{H}=-\frac{\alpha H^r}{2(1+\beta H^s)}
\label{dhdt}
\end{equation}
where again $\alpha$, $\beta$, $r$, and $s$ are constants related to those of (\ref{van1}) through the relations
\begin{equation}
r=2 n \; ; \; \; \; s=2 m \; ; \; \; \; \alpha=3^n \gamma \; ; \; \; \; and \; \; \; \beta=3^m \delta
\label{const}
\end{equation}

The solution of equation(\ref{dhdt}) indicates the evolution of the Hubble parameter with time. This equation is integrated to give
\scriptsize
\begin{equation}
\frac{\alpha}{2}(t_0-t)=\frac{1}{1-r}\left({H_0}^{1-r}-{H}^{1-r}\right) + \frac{\beta}{s-r+1}\left({H_0}^{s-r+1}-H^{s-r+1}\right)
\label{solh}
\end{equation}
\normalsize
However, our model is not describing a pure dark energy, in which case the values of the parameters are totally free and are not constrained. Instead, our model able to follow the smooth phase transition which takes place through the universe evolution. This restricts the parameter values to obey the limits satisfied by observations so that for early time we must have the equation of state of the perfect fluid.

Accordingly, we first have to study the asymptotic behavior of $\dot{H}$. From one side, this indicates to what extent our model agrees the real evolution of the universe, and from the other side may also fixes the limits of some of our model parameters.  Table \ref{tab:ass} shows this behavior for the different ranges of of the parameters $r$ and $s$.

\begin{table}[t]
\small
	\centering
	\caption{\footnotesize{Asymptotic behavior of the equation of state}} 
		\begin{tabular}{@{}cccc@{}}
		\hline
		 Case &ranges of $r$ and $s$ & large $H$ & small $H$  \vspace*{1pt} \\  \hline \vspace{-7pt} \\
		$I$ & $r \succ 0$ $\&$ $s \succ 0$ &  $\dot{H}{\rightarrow}\frac{-\alpha}{2\beta} H^{r-s}$  &  $\dot{H}{\rightarrow}\frac{-\alpha}{2} H^r$ \vspace*{1pt} \\
		$II$ & $r \succ 0$ $\&$ $s \prec 0$ &  $\dot{H}{\rightarrow}\frac{-\alpha}{2} H^{r}$  &  $\dot{H}{\rightarrow}\frac{-\alpha}{2\beta} H^{r-s}$ \vspace*{1pt} \\
		$III$ & $r \prec 0$ $\&$ $s \succ 0$ &  $\dot{H}{\rightarrow}\frac{-\alpha}{2\beta} H^{r-s}$  &  $\dot{H}{\rightarrow}\frac{-\alpha}{2} H^r$ \vspace*{1pt} \\
		$IV$ & $r \prec 0$ $\&$ $s \prec 0$ &  $\dot{H}{\rightarrow}\frac{-\alpha}{2} H^{r}$  &  $\dot{H}{\rightarrow}\frac{-\alpha}{2\beta} H^{r-s}$ \vspace*{1pt} \\	\hline		
		\end{tabular}
	\label{tab:ass}
\end{table}

One can see from the table  that the values of negative $r$ must be precluded in our calculations as a result of the asymptotic behavior of the $\dot{H}$ function. This is because for asymptotically perfect fluid behavior of our universe, the $\dot{H}$ function at large $H$ forces $r$ to be positive if $s$ is negative, and forces $(r-s)$ to be positive if $s$ is positive, which does not satisfy constraints for the pre-assumptions.

On the other hand, the asymptotic behavior of $\dot{H}$ for the two cases of positive $r$ can attain the perfect fluid behavior at large $H$ if $r-s=2$ and $\alpha / \beta =3$ for the case of $+ve \; s$ and if $r=2$ and $\alpha=3$ for the case of $-ve \; s$. These two cases will be studied in details.

\section{Analyzing the Model} \label{anal}
\begin{figure}
	\centering
		\includegraphics[width=18cm, height=25cm] {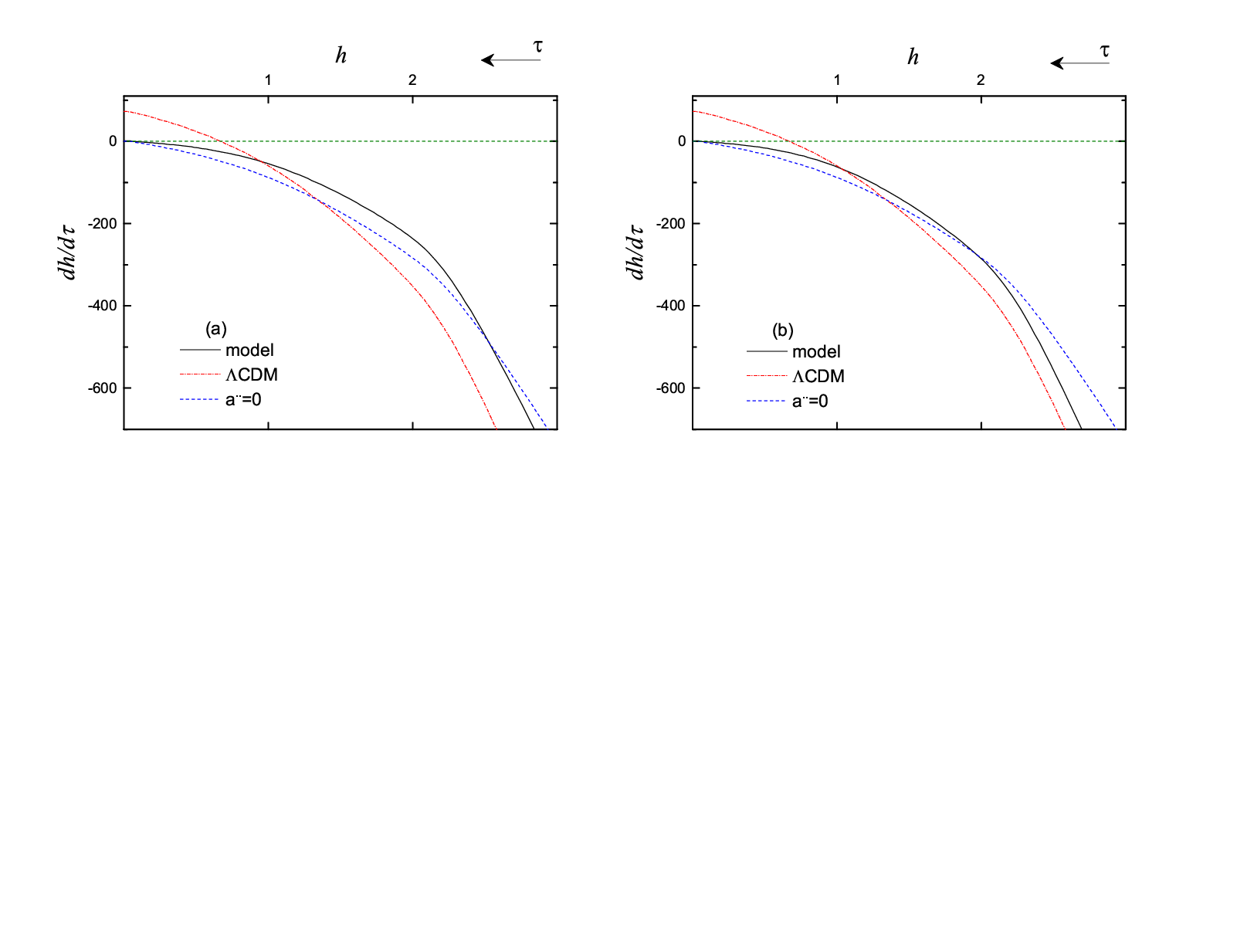}
  	\vspace{-8cm}
	\caption{Phase diagram of our model (solid black line), case $(a)$ for $+ve \; s$ and case $(b)$ for $-ve \; s$. Shown also $\Lambda CDM$     curve (dashed red line) and the zero acceleration curve (dotted blue line) for comparison.}
	\label {fig:phase}
\end{figure}

The phase space analysis method enables information about the behavior of the cosmological model and its classes of solutions. In Fig.\ref{fig:phase} we show a phase diagram, $H-\dot{H}$ diagram, for our model agaist $\Lambda CDM$ model, where $h=\frac{H}{H_0}$ and $\tau=\frac{t}{t_0}$. In the Fig. we also plot the zero acceleration curve for which
\begin{equation}
\ddot{a}=\dot{H}+H^2=0
\end{equation}
so that
\begin{equation}
\dot{H}=-H^2
\end{equation}

Due to \cite{aawad}, since the function $F(h)=dh/d\tau$ is continuous and differentiable and since we have future fixed point, our model has a unique solution for times $\tau \succ 0$ and has no future finite time singularities. instead, it evolves to a de Sitter space. 

We can see from the Fig. that deceleration parameter changes sign from $+ve$ to $-ve$ as the universe evolved from matter dominated era to dark energy dominated era. We can also see that our model expects matter-DE equality time to be earlier than expected by $\Lambda CDM$ model.

%
%
\section{Case I: $s$ is positive} \label{case1}
This is the first case that match the dark fluid dynamics. The asymptotic behavior of the EoS at large $H$, early times, has the form $\dot{H}{\rightarrow} \frac{-\alpha}{2\beta} H^{r-s}$, so that for $r-s=2$ and $\alpha / \beta =3$ we have the perfect fluid EoS at early times. 
Let's now examine the constraints that can be obtained by applying the causality condition to this model.

\subsection{Causality Constraints} \label{causp}
Spherical sound waves are lunched as a result of the overpressure due to the initial overdensity in the DM and gas. These oscillations have a speed which is an indication to the velocity by which such perturbations are transmitted. Accordingly, it must not exceed the speed of light, which is a condition of causality. 

For a barotropic cosmic fluid, the speed of these sound waves is defined by the relation
\begin{equation}
{c_{s}}^2 = \frac{dp}{d \rho}
\label{causcon}
\end{equation}
As a result, it is an imprint for the barotropic models to have a value of $d p/d \rho$ constrained by limits on the sound speed \cite{lin}. This is because $d p/d \rho$ must satisfy the condition
\begin{equation}
\frac{dp}{d \rho} \geq 0
\label{con1}
\end{equation}
to insure stability in perturbation growth, so that we must have ${c_{s}}^2 \geq 0$. On the other hand, causality constrains ${c_{s}}^2$ not exceed the speed of light \cite{ell}, \cite{bab}, so that ${c_{s}}^2 \leq 1$. This fact distinguishes the dynamical behavior of barotropic fluids from quintessence models which describe an imperfect fluid and generically have $\frac{dp}{d \rho} \prec 0$ \cite{cald}. As a prototype example for barotropic fluids is the Chaplygin gas \cite{kam} and the generalized Chaplygin gas models \cite{bent} which have an EoS 
\begin{equation}
p=-\frac{A}{\rho^\alpha}
\label{chga}
\end{equation}
where $A$ and $\alpha$ are $+ve$ constants. For these models we have 
\begin{equation}
{c_{s}}^2=\frac{\alpha A}{\rho^{\alpha+1}}
\label{chga1}
\end{equation}
which ensures ${c_{s}}^2 \geq 0$ as $A$ and $\alpha$ are $+ve$.

 Accordingly, applying constraints for barotropic fluids, one may get an indication for the available ranges for the values of the free parameters in the EoS. Now using (\ref{van1}), we get 
\begin{equation}
{c_{s}}^2=-1+\frac{\left[\left(n-m\right)\delta \rho^m+n\right]\gamma \rho^{n-1}}{\left(1+\delta \rho^m\right)^2} \leq 1
\label{cas1}
\end{equation}
\normalsize

Now for our case where $n$ and $m$ are $+ve$, relation (\ref{cas1}) for large $\rho$ gives the condition
\begin{equation}
\left(n-m\right)\frac{\gamma}{2\delta} \rho^{n-m-1} \leq 1
\label{cas6}
\end{equation}
This is fully satisfied for 
\begin{equation}
n\leq m+1
\label{cas7}
\end{equation}
While for small $\rho$, the condition will be
\begin{equation}
\frac{\left(n-m\right)}{2} \gamma \delta \rho^{n+m-1} + \frac{n}{2} \gamma \rho^{n-1} \leq 1
\label{cas8}
\end{equation}
This is satisfied if
\begin{equation}
n+m\geq 1 \;\;\; and \;\;\; n\geq 1
\label{cas9}
\end{equation}
Of course, this last condition is automatically satisfied if the the condition in (\ref{cas7}) is satisfied. 
Now Combining our conditions, (\ref{cas7}) and (\ref{cas9}), we get to the conclusion that for this case we must consider
\begin{equation}
1-m \leq n \leq 1+m
\label{cas10}
\end{equation}
Which means
\begin{equation}
2(1-s) \leq r-s \leq 2
\label{cas11}
\end{equation}
In agreement with the conditions required by asymptotic behavior of the EoS at early time for this case.
In fact, for the general values of $\rho$ we also have a large dependence on $\gamma$ and $\delta$. Accordingly, we have to make a check using the obtained optimized values for the parameters.

\subsection{The dynamics of the universe} \label{secdyp}
In this section we study the ability of our model to produce the physical quantities that can be really measured. This, of course, depends on the compatibility of the results of our model with the astrophysical data that are indeed observed. 

\subsubsection{The Luminosity Distance} \label{lum1p}
One of the most important observed data is that of the Hubble diagram of type SNe Ia supernovae, which is a plot of the luminosity distance $D_L$, that determines the flux of the source, as a function of the red shift. We will use these data to constrain the parameters in our model. Due to the asymptotic and causality constraints, we have now only two out of four parameters that are free and needed to be optimized due to observations.

\begin{figure}[h]
	\centering
		\includegraphics [width=10cm,height=8cm]{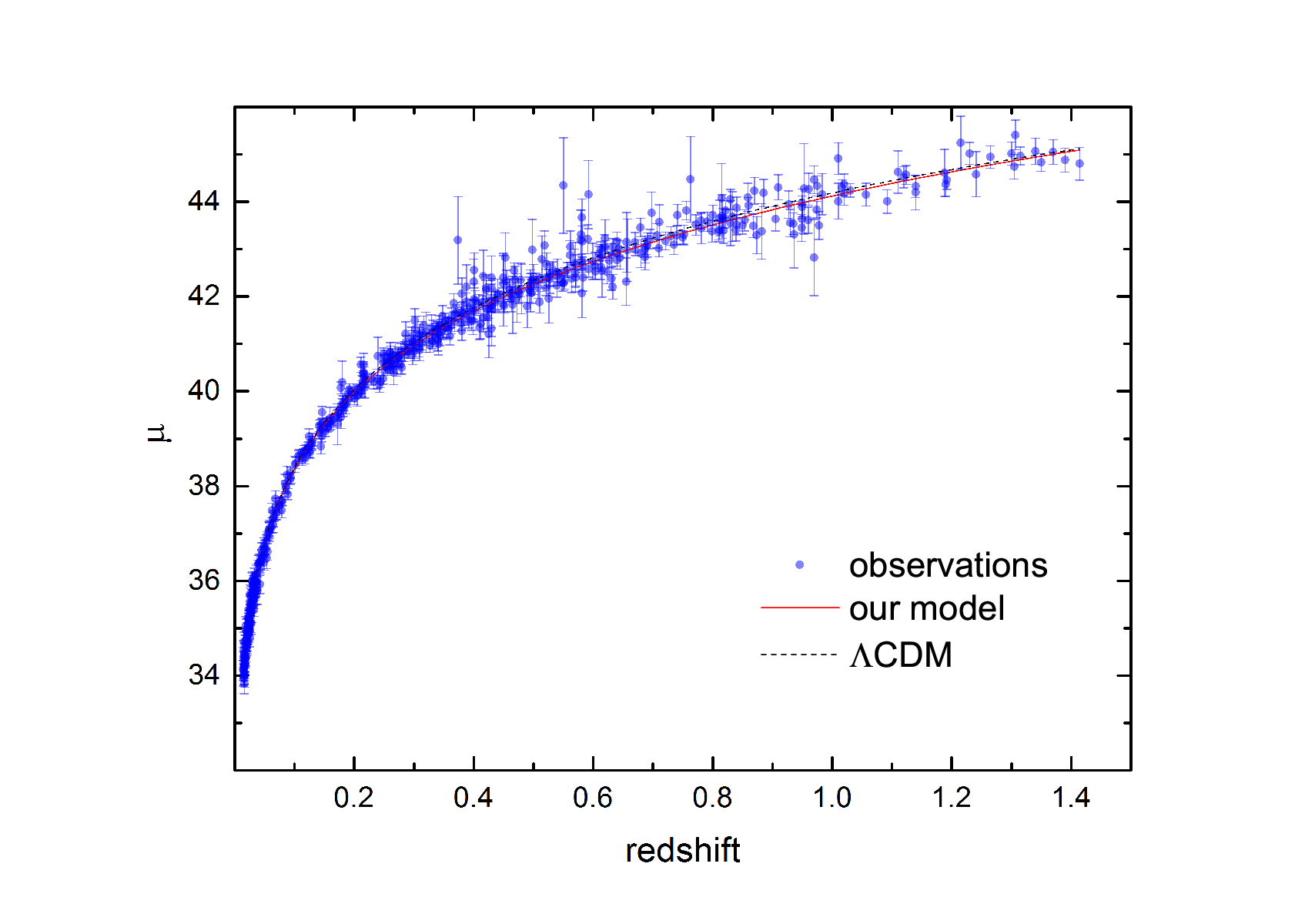}
		\vspace{-.6cm}
	\caption{\footnotesize{Model versus observations for SNe Ia supernovae for of $+ve  \, s$ case. Results from $\Lambda CDM$ model, dashed line, are also shown for comparison.}}
	\label{fig:lmnstyp}
\end{figure}

To satisfy the causality and asymptotic behavior constraints, we choose $(r-s)$ as $2$ while $(\alpha/\beta)$ has to be $3$, so that we can attain the perfect fluid behavior of $\dot{H}$ at large $H$. Let's now start with our model by using the relation (\ref{dhdt}) for $(r-s)=2$  in (\ref{rodot}), so that we have
\begin{equation}
\dot\rho=-3 H \frac{\alpha H^{s+2}}{1+\beta H^s}
\label{rohdotp}
\end{equation}
which in turn gives
\begin{equation}
\frac{dH}{da}=-\frac{1}{2 a} \frac{\alpha H^{s+1}}{1+\beta H^s}
\label{dhdap}
\end{equation}
This has the solution
\begin{equation}
H(a)=a^{-\alpha/{2 \beta}} \exp\left[{\frac{1}{s} W\left(\frac{1}{\beta} a^{\frac{\alpha s}{2 \beta}} e^{C \frac{\alpha s}{2 \beta}}\right)} e^{\frac{-\alpha}{{2 \beta}} C} \right]
\label{hofap}
\end{equation}
\normalsize
The function $W(x)$ is the Lambert W-function. To fix the constant C we consider that at some time where $t=t_0$ we have $a=a_0$ and $H=H_0$, this gives
\begin{equation}
C=- \ln\left(a_0 {H_0}^{{2 \beta}/\alpha}\right) + \frac{2}{\alpha s} {H_0}^{-s}
\label{cofhp}
\end{equation}
So that
\scriptsize
\begin{equation}
H(a)=\left({\frac{a_0}{a}}\right)^{\frac{\alpha}{2 \beta}} H_0 \, e^{\frac{-{H_0}^{-s}}{\beta s}} \exp\left[{\frac{1}{s}} W\left(\left(\frac{a_0}{a}\right)^{\frac{-\alpha s}{2 \beta}} \frac{H_0^{-s}}{\beta} \, e^{\frac{H_0^{-s}}{\beta}}\right) \right] 
\label{hofap1}
\end{equation}
\normalsize
Accordingly
\scriptsize
\begin{equation}
H(z)=\left(z+1\right)^{\frac{\alpha}{2 \beta}} H_0 \, e^{\frac{{-H_0}^{-s}}{\beta s}} \exp\left[{\frac{1}{s} W\left(\left(z+1\right)^{\frac{-\alpha s}{2 \beta}} \frac{{H_0}^{-s}}{\beta} \, e^{\frac{{H_0}^{-s}}\beta}\right)} \right] 
\label{hofzp}
\end{equation}
\normalsize
the luminosity distance $D_L$ is then given by
\begin{equation}
D_L(z)=(1+z) D_p(z)
\label{dl}
\end{equation}
where $D_p(z)$ is the proper distance given by
\begin{equation}
D_p(z)={\int_0}^z \frac{1}{H(\zeta)} d\zeta
\label{dp}
\end{equation}
Accordingly
\scriptsize
\begin{eqnarray}
D_p(z)=\frac{e^{\frac{{H_0}^{-s}}{\beta s}}}{H_0} \cdot \qquad \qquad \qquad \qquad \qquad \qquad \qquad \qquad \qquad \qquad\nonumber \\
       \qquad {\int_0}^z {\left(\zeta+1\right)}^{\frac{-\alpha}{{2 \beta}}} \exp\left[{\frac{-1}{s} W\left(\left(\zeta+1\right)^{\frac{-\alpha s}{2 \beta}} \frac{{H_0}^{-s}}{\beta} \, e^{\frac{{H_0}^{-s}}{\beta}}\right)} \right] d\zeta 
\label{dpofzp1}
\end{eqnarray}
\normalsize
The integration of the above equation can not be solved analytically. Accordingly, we used numerical methods to solve the integral. This enabled exact solution for the relation controlling the dependence of the luminosity distance on the redshift. The comparison of our model against supernovae data relies on the $\chi^2-$statistics where
\begin{equation}
\chi^2={\sum}_{i}{\frac{\left(\mu_{th}\left(z_i\right)-\mu_{obs}\left(z_i\right)\right)^2}{{\sigma_i}^2}}
\label{chisqr}
\end{equation}
where $\mu$ is the distance modulus. In Fig.\ref{fig:lmnstyp} we plot our model against data set from Conley et al. \cite{con}. Data set are available at Supernova Cosmology Project. Results from $\Lambda CDM$ model \cite{eaa} are also shown in the Fig. for comparison.

In terms of distance modulus, data points are 
\begin{equation}
\mu_{obs}(z)=m(z)-M
\label{distmod}
\end{equation}
where $M$ is the absolute magnitude which is assumed to be constant for a standard candle like SNe Ia, while $m$ is the apparent magnitude given by
\begin{equation}
m(z)=M+5 log_{10} \left(\frac{D_L}{MPc}\right)+25
\label{appmag}
\end{equation}

\begin{figure}[h]
    	\centering
			\vspace{.5cm}
		\includegraphics [width=8cm,height=6cm]{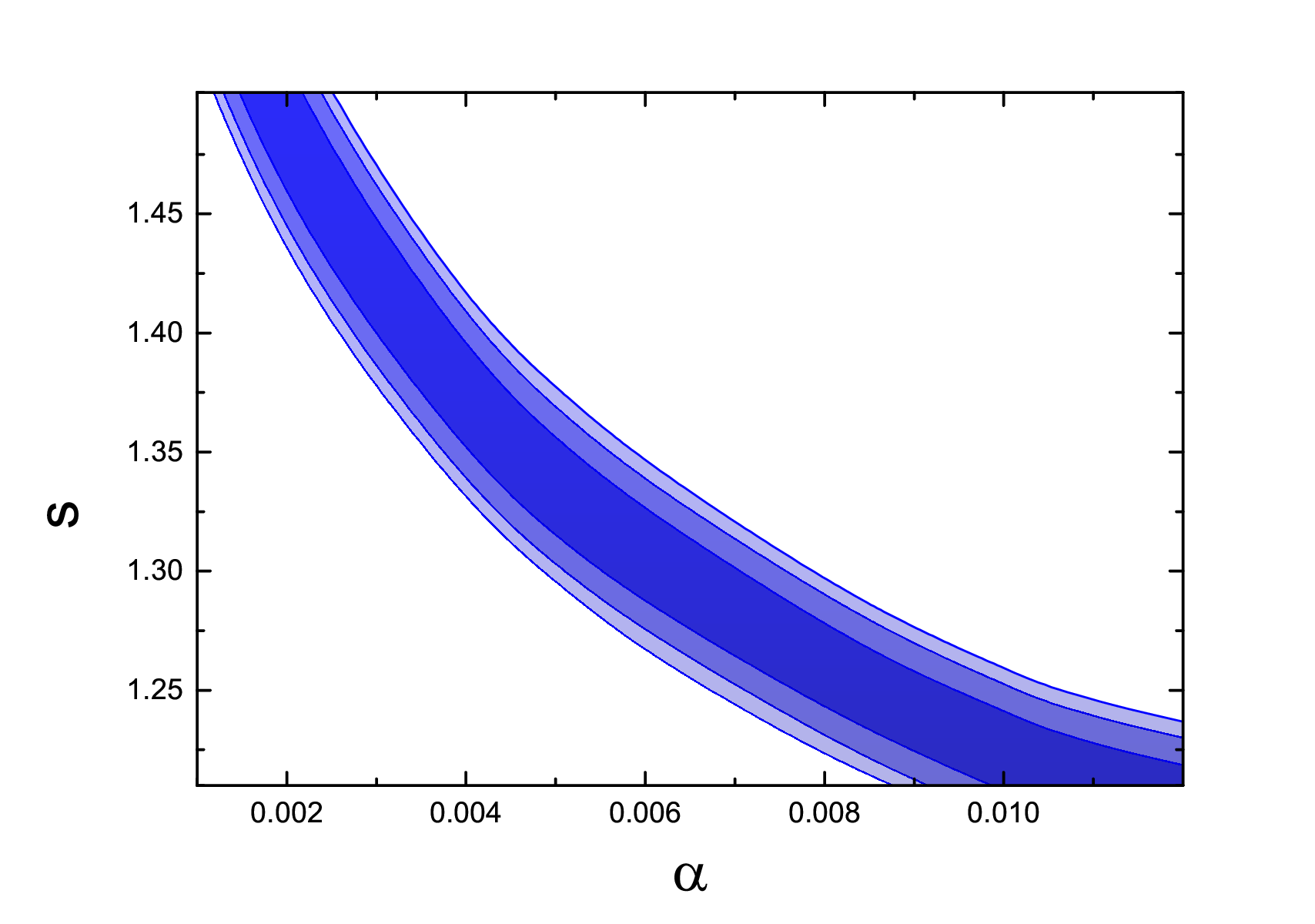}
	  \vspace{-.5cm}
	\caption{\footnotesize{Confidence regions in parameters space for $1 \sigma$, $2 \sigma$, and $3 \sigma$.}}
	\label{fig:cont}
\end{figure}

Using $H_0=67.8 Km/s/MPc$ due to Plank $+$ WMAP $+$ BAO measurements \cite{plk}, we obtained a best fit of our model at $\alpha=0.004$ and $s=1.369$ for which ${\chi^2}_{red}=1.020$. This result has a $Q-$value $=0.34$ which shows that data are consistent with the model. Fig.\ref{fig:cont} shows the confidence regions in parameters space.

\begin{figure}[h]
	\centering
		\includegraphics[width=9cm, height=7cm]{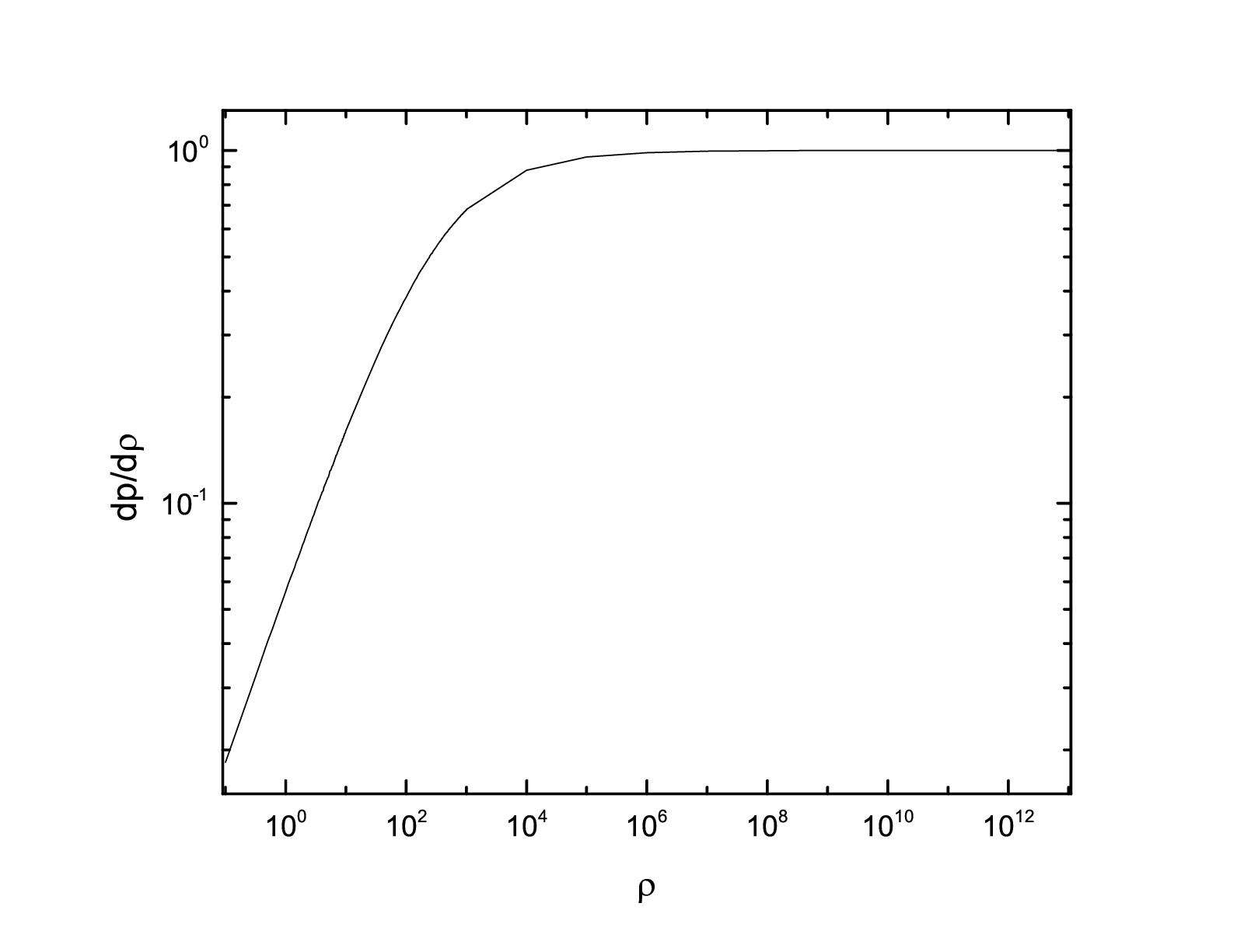}
	
	\vspace{-.4cm} 
	\caption{\footnotesize{Causality test using best parameters for $+ve \, s$}}
	\label{fig:pstvs}
\end{figure}

Having the full set of parameters, let's test the causality condition using these parameters. A plot of causality relation (\ref{cas1}) using our parameters for this case is shown in Fig(\ref{fig:pstvs}). We can see from the figure that the parameters satisfy the conditions for ${c_s}^2$ throughout the whole stages of the evolution of the universe.

\subsubsection{The Energy Density Evolution} \label{enrgp}
Let us see how the energy will evolve with time due to the model in a homogenous and isotropic universe. Relations (\ref{cons}) and (\ref{van1}) give
\begin{equation}
d\rho=\frac{1}{a} \frac{-3 \gamma \rho^n}{1+\delta \rho^m} da
\label{roh1}
\end{equation}
To simplify our way of solution to this equation let's use our parameters. In this case we have $n=m+1$, so that the solution of relation (\ref{roh1}) will take the form
\scriptsize
\begin{equation}
\rho(a)=\rho_0 \left({\frac{a}{a_0}}\right)^{-3 \gamma / \delta} \exp{\left(\frac{W(x)}{m}-\frac{1}{m \delta {\rho_0}^m}\right)}
\label{roh8}
\end{equation}
\normalsize
Where
\scriptsize
\begin{equation}
x=\frac{1}{\delta {\rho_0}^m} e^{\frac{1}{\delta {\rho_0}^m}} \left({\frac{a}{a_0}}\right)^{3 \gamma m / \delta}
\label{roh9}
\end{equation}
\normalsize
To study the properties of relation (\ref{roh8}) at early times we make use of the property
 \begin{equation}
\exp\left(n W(x)\right)=\left(\frac{x}{W(x)}\right)^n
\label{roh322}
\end{equation}
then \begin{equation}
\exp\left(\frac{W(x)}{m}\right)=\left(\frac{x}{W(x)}\right)^{1/m}
\label{roh333}
\end{equation} 
Now, in the past, where $a\prec\prec a_0$, so that $x \prec\prec 1$ due to (\ref{roh9}) as $m$ is $+ve$, Taylor expansion of $W$-function can be approximated to $W(x)\approx x$, so that the exponential term of the $W$-function converges to $1$, then the energy density for that time will take the asymptotic form
\begin{equation}
\rho(a) \rightarrow A \left({\frac{a}{a_0}}\right)^{-3}
\label{roh10}
\end{equation}
as we have $\gamma / \delta =1$ for this case, where $A = \rho_0 e^{\frac{-1}{\delta m {\rho_0}^m}}$ is a constant. Accordingly, the fluid of the model behaves as a matter in the earlier times.

On the other hand, for $a \succ\succ a_0$, $x \succ\succ 1$, the energy density function due to relation (\ref{roh333}) will tend to
\begin{equation}
\rho(a)=\left({\delta x}\right)^{-1/m} \left(\frac{x}{W}\right)^{1/m}
\label{roh11}
\end{equation}
Accordingly
\begin{equation}
\rho(a)= \left(\frac{1}{\delta W(x)}\right)^{1/m}
\label{roh20}
\end{equation}
Using the value of $m$ we can see that this function is a very slowly decreasing function of $x$ for $x \succ 1$, and tends to be nearly constant at $x \succ\succ 1$, i.e. at $a \succ\succ a_0$, a signature for a cosmological constant towards the future.

The equation of state parameter $\omega$ for this case, where $m=n-1$, will be given, from (\ref{par1}), by
\begin{equation}
\omega=-1+\frac{\gamma}{\rho^{-m}+\delta}
      =-1+\frac{1}{\left(\rho\right)^{-m} / \gamma+\delta / \gamma}
\label{roh12}
\end{equation}
Using our parameters for this case where $\delta / \gamma=1$, we finally get
\begin{equation}
\omega =-1+\frac{1}{\left(\rho\right)^{-m} / \gamma+1}    
\label{roh22}
\end{equation}
Indicating that for earlier times, where we have large values of $\rho$, $\omega \rightarrow 0$, a signature for matter era, while changes to $-1$ gradually as the matter content of the universe diluted.

\subsubsection{Deceleration Parameter and the Universe Acceleration} \label{accelrp}
The acceleration of the universe is related to a dimensionless parameter known as the deceleration parameter $q$ defined as
\begin{equation}
q = -\frac{\ddot{a}}{a} \frac{1}{H^2}
\label{dec1}
\end{equation}
Since you have to have $\frac{\ddot{a}}{a} \succ 0$ to get an accelerating universe, the deceleration parameter will then measure if the expansion of the universe is accelerating or decelerating. If $q \prec 0$, the expansion is accelerating, while if $q \succ 0$ it is decelerating. Now as 
\begin{equation}
\frac{\ddot{a}}{a}=H^2+\dot{H}
\label{dec2}
\end{equation}
Then 
\begin{equation}
q = -1 - \frac{\dot{H}}{H^2}
\label{dec3}
\end{equation}
Now using (\ref{dhdt}), one gets the deceleration parameter for our model to be
\begin{equation}
q=-1 + \frac{\alpha H^{r-2}}{2(1+\beta H^s)}
\label{dec4}
\end{equation}
Using $r=s+2$, relation (\ref{dec4}) gives
\begin{equation}
q=-1 + \frac{\alpha}{2(H^{-s}+\beta)}
\label{dec6}
\end{equation}
Using the values of our parameters with $H_0=67.8$, the present day deceleration parameter due to this model case has the value
\begin{equation}
q_0 = -0.550
\end{equation}
Relation (\ref{dec6}) also shows that deceleration parameter evolves towards more negative values as the Hubble parameter decreases with time.

%
%
\section{Case II: $s$ is negative} \label{case2}
For this case $\dot{H}\rightarrow \frac{-\alpha}{2} H^{r}$ asymptotically at large $H$, which means that we can attain the perfect fluid EoS at early time if $r=2$ and $\alpha=3$. However, a first step again is the examination of the causality constraints for this case of the model.

\subsection{Causality Constraints} \label{caus2}
In this case $m$ is $-ve$, so that relation (\ref{cas1}) at large values of $\rho$ tends to
\begin{equation}
\frac{1}{2} n \gamma \rho^{n-1}\leq1
\label{cas2}
\end{equation}
Which is satisfied if
\begin{equation}
0 \prec n \leq1 
\label{cas3}
\end{equation}
while for small $\rho$ we have
\begin{equation}
\left(n-\frac{m}{2}\right)\frac{\gamma}{\delta} \rho^{n-m-1}\leq 1
\label{cas4}
\end{equation}
As $m$ is $-ve$, this is always satisfied under the condition
\begin{equation}
n\geq 1
\label{cas4d}
\end{equation}
 Combining (\ref{cas3}) and (\ref{cas4d}) we get to the conclusion that in this case we must consider
\begin{equation}
n=1 
\label{cas5}
\end{equation}
which means
\begin{equation}
r=2
\label{cas55}
\end{equation}
in full agreement with the asymptotic behavior of the EoS at early time for this case.

\subsection{The dynamics of the universe} \label{secdyn}
We'll now study the physical quantities representing the dynamics of our universe due to this case of the model and comparing our results with observations.

\subsubsection{The Luminosity Distance} \label{subln}
\begin{figure}[htbp]
	\centering		\includegraphics[width=10cm,height=8cm]{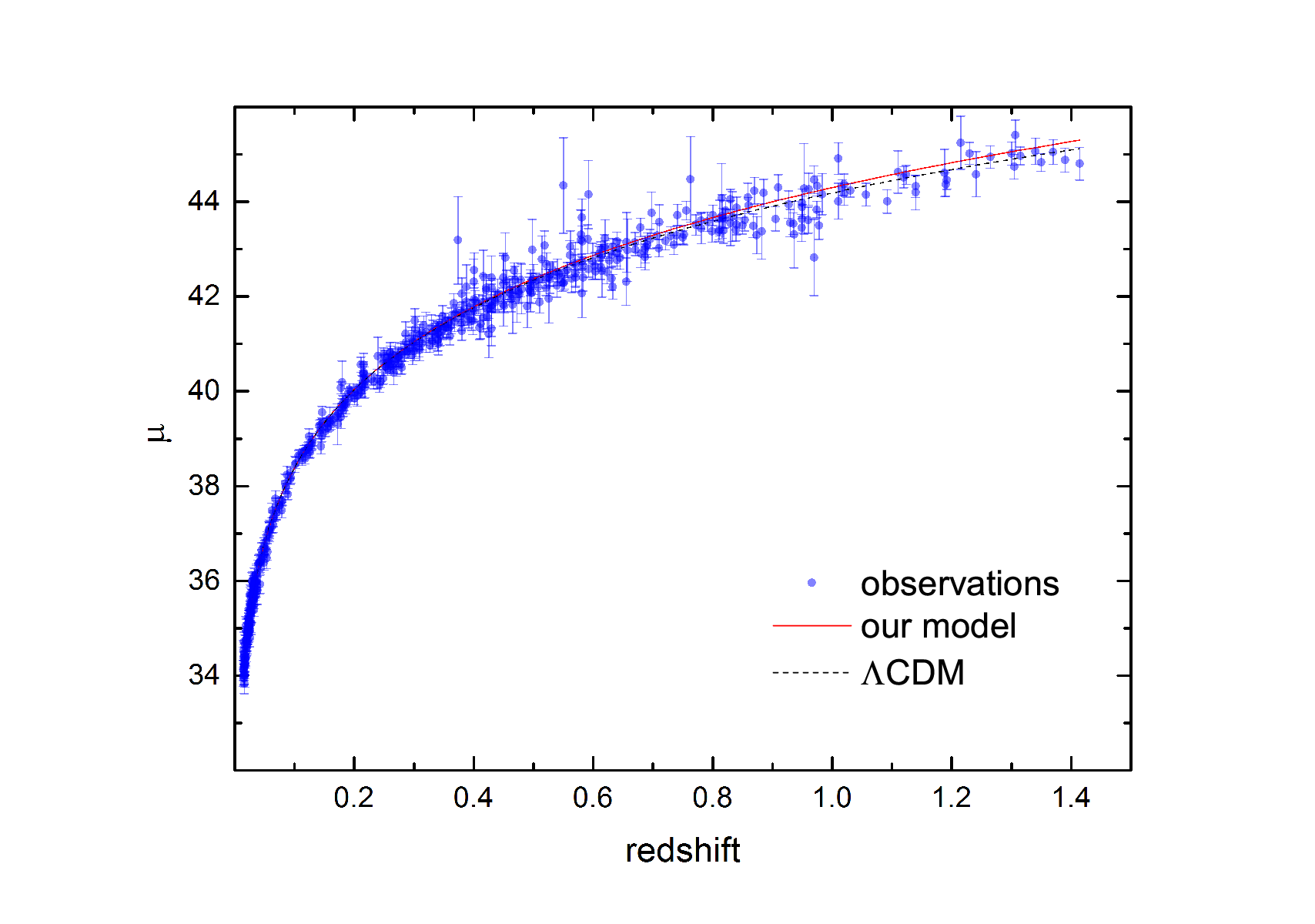}
		\vspace{-.6cm}
	\caption{\footnotesize{model versus observations for SNe Ia supernovae for $-ve \, s$. Results from $\Lambda CDM$ model, dashed line, are shown for comparison.}}
	\label{fig:lmnsty}
\end{figure}

We could now constrain two parameters out of four through asymptotic behavior of the model and the causality condition, where $r$ has to be $2$ while $\alpha$ has to be $3$. Let's then calculate the luminosity distance function $D_L(z)$ and use the observed data of this function to constrain the remaining two parameters of the model. We start by using relation (\ref{dhdt}) for $r=2$ in (\ref{rodot}), so that we have
\begin{equation}
\dot\rho=-3 H \frac{\alpha H^2}{1+\beta H^s}
\label{van3}
\end{equation}
which in turn gives
\begin{equation}
\frac{dH}{da}=-\frac{1}{a} \frac{\alpha H}{1+\beta H^s}
\label{dhda}
\end{equation}
This has the solution
\scriptsize
\begin{equation}
H(a)=a^{-\alpha/2} \exp\left[{-\frac{1}{s} W\left(\beta a^{-\alpha s/2} e^{-C \alpha s/2}\right)} - \frac{C \alpha}{2} \right]
\label{hofa}
\end{equation}
\normalsize
And let's assume that at some time $t=t_0$ we have $a=a_0$ and $H=H_0$. This gives
\begin{equation}
C=- \ln\left(a_0 {H_0}^{2/\alpha}\right) - \frac{2}{\alpha s} \beta {H_0}^s
\label{cofh}
\end{equation}
So that
\scriptsize
\begin{equation}
H(a)=\left({\frac{a_0}{a}}\right)^{\frac{\alpha}{2}} H_0 \exp\left(\frac{\beta {H_0}^s}{s}\right) \exp\left[{-\frac{1}{s} W\left(\beta \left(\frac{a_0}{a}\right)^{\frac{\alpha s}{2}} {H_0}^s e^{\beta {H_0}^s}\right)} \right]
\label{hofaa}
\end{equation}
\normalsize
In turn, we can write $H(z)$ as
\scriptsize
\begin{equation}
H(z)=\left(z+1\right)^{\alpha/2} H_0 \exp\left(\frac{\beta {H_0}^s}{s}\right) \exp\left[{-\frac{1}{s} W\left(\beta \left(z+1\right)^{\alpha s/2} {H_0}^s e^{\beta {H_0}^s}\right)} \right]
\label{hofz}
\end{equation}
\normalsize
The luminosity distance $D_L$, Eqs (\ref{dl}), is then calculated using the proper distance which due to (\ref{dp}) will be given by
\scriptsize
\begin{eqnarray}
D_p(z)=\frac{1}{H_0} \exp\left(-\frac{\beta {H_0}^s}{s}\right) \cdot \qquad \qquad \qquad \qquad \qquad \qquad \qquad \qquad \qquad \qquad   \nonumber \\
{\int_0}^z {\left(\zeta+1\right)}^{-\alpha/2} \exp\left[{\frac{1}{s} W\left(\beta \left(\zeta+1\right)^{\alpha s/2} {H_0}^s e^{\beta {H_0}^s}\right)} \right] d\zeta \qquad 
\label{dpofz}
\end{eqnarray}
\normalsize
Again, the integration of relation (\ref{dpofz}) can not be solved analytically so that we use numerical methods. The apparent magnitude of Conley et al. data set, \cite{con}, is plotted in Fig.\ref{fig:lmnsty} together with our results. Results from $\Lambda CDM$ model are also shown for comparison.

applying tha $\chi^2-$statistics, relation (\ref{chisqr}), We obtained a best fit of this model case at $s=-2.1$ and $\beta=1.64 \times 10^4$ for which ${\chi^2}_{red}=1.23$. However, the $Q-$value of this result is $1.05\times10^{-4}$. This $Q-$value in fact tells us that data are not consistent to be drawn from this model case, and so this case will be ruled out.

\section{Conclusions} \label{seconc}
A barotropic fluid model is considered which initially can be described as a correction to the dark energy regime by an an EoS that has two power laws asymptotes. This enables interpolation between the DM and DE equations of state, and so guarantees a smooth phase transition from matter era to DE era. Accordingly, our model is a dark fluid model provides a unification for the two dark sectors in one EoS that has the advantage of the general asymptotic solution. This means that it can describe dark matter at early times while describing dark energy at late times, while adopts a general EoS for DE which enables the cosmological constant as a special case.

Studying the asymptotic behavior of the equation of state constrains the parameters of the equation and restricts the model. It also clarifies the cases that can be physically acceptable as dark fluid which are found to be two cases. On the other hand, studying the phase diagram of the model for each of these two cases shows that our model has a unique solution for times $\tau \succ 0$, evolves to a de Sitter space and has no future finite time singularities.

Causality condition further constrains the parameters of the model, so that at the end, only two out of four parameters are free. These two parameters are constrained using the Hubble diagram of SNe Ia supernovae, where we present an exact solution for the red shift dependence of the luminosity distance in each model, or model case, avoiding the approximate formulas that are accurate only at small red shifts. 

The $\chi^2-$statistics shows that the $+ve \; s$ case can fit the data well. Accordingly, We studied the energy density evolution for this model case which ensures the smooth transition between the two dark sectors. Furthermore, we examined the evolution of the EoS parameter $\omega$ that again ensures its two extremes of $0$ and $-1$ for early and late times. We also Calculated the deceleration parameter $q$ and examined its evolution. Calculations resulted in a negative $q$ with the present day value of $-0.550$, while assures evolution towards more negativity to the future as the Hubble parameter decreases, a clear evidence for accelerated expansion in agreement with observations. On the other hand, the $-ve \; s$ case failed to fit the data and so will be ruled out.

Due to the above mentioned results, our model with a $+ve \; s$ parameter can be considered as a dark fluid model unifying the two dark sectors of the universe and succeeded in describing the evolution of the universe.

\section{Aknowledgment} \label{akn}
The author would like to thank Dr. Adel Awad for valuable discussions and guidance.

\end{document}